\documentclass[twocolumn,superscriptaddress,pre,showpacs,preprintnumbers,amsmath,amssymb]{revtex4}
\usepackage{CJK}
\usepackage{amsmath}
\usepackage{amssymb}
\usepackage{dcolumn}
\usepackage{graphicx}
\usepackage{bm}

\begin{document}
\begin{CJK}{GB}{gbsn} 
\title{$[N]pT$ ensemble and finite-size scaling study of the GEM-4 critical isostructural transition}
\author{Kai Zhang}  
\affiliation{Department of Chemistry, Duke University, Durham, North
Carolina, 27708, USA}
\author{Patrick Charbonneau}
\affiliation{Department of Chemistry, Duke University, Durham, North
Carolina, 27708, USA}
\affiliation{Department of Physics, Duke University, Durham, North
Carolina, 27708, USA}
\date{\today}

\begin{abstract}
First-order transitions of system where both lattice site occupancy and lattice spacing fluctuate, such as cluster crystals, cannot be efficiently studied by traditional simulation methods, which necessarily fix one of these two degrees of freedom. The difficulty, however, can be surmounted by the generalized $[N]pT$ ensemble [J. Chem. Phys. \textbf{136}, 214106 (2012)]. Here we show that histogram reweighting and the
$[N]pT$ ensemble can be used to study an isostructural transition between cluster crystals of different occupancy in the generalized exponential model of index 4 (GEM-4).  Extending this scheme to finite-size scaling studies also allows to accurately determine the critical point parameters and to verify that it belongs to the Ising universality class.
\end{abstract}
\maketitle
\end{CJK}

Models with steeply growing but soft-core repulsions were first introduced as schematics for soft matter systems, such as dendrimers and micellar crystals~\cite{likos:2001,mladek:2008,lenz:2011}. In contrast to models with purely repulsive hard-core interactions, which simply crystallize, at high densities these models form cluster crystals in which each lattice site is multiply occupied. Clustering dramatically affects the materials properties of the crystal phase, contributing to both these model's response functions and elastic properties~\cite{mladek:2008,zhang:2012}. One of the canonical cluster-crystal formers, the generalized exponential model of index $4$ (GEM-4)~\cite{mladek:2006} with a pair interaction potential
\begin{equation}
\phi(r) = \epsilon \exp[-(r/\sigma)^4],
\end{equation}
where $\epsilon$ and $\sigma$ set the units of energy and length respectively, was also found to exhibit low-temperature first-order isostructural transitions between cluster phases with different integer occupancy, e.g., face-centered cubic (FCC) solids with double (FCC2) and triple (FCC3) occupancy~\cite{zhang:2010b}. Each of these isostructural transitions was found to terminate at a critical temperature $T_c$, above which the lattice difference between the two phases vanishes and particles are randomly distributed over the lattice sites.

Although rather rare, isostructural transitions between singly occupied FCC solids with different lattice spacing were first observed in the cerium phase diagram over thirty years ago~\cite{young:1977,young:1979}, and then in models with short-range attractive interactions~\cite{young:1980,bolhuis:1994}.  No direct study of their critical properties has  been reported, but these properties are expected to be fairly standard, except in two dimensions, where coupling with a hexatic transition may affect the critical behavior~\cite{chou:1996}.  In the case of cluster crystals, however, the interplay between lattice occupancy and lattice spacing results in additional sources of fluctuation that could impact the critical behavior, just like they affect these system's mechanical response. From a numerical simulation point of view, clustering also prevents the successful application of traditional simulation methods. In this Brief Report, we present a finite-size scaling study of the FCC2-FCC3 critical point in the GEM-4 by specially considering the lattice occupancy fluctuation at equilibrium. The approach confirms the Ising universality class of the transition and finds a minimal role for coupling. We expect that the approach could be extended to other systems where lattice occupancy plays a central role, such as microphase formers~\cite{seul:1995}, crystals with large number of vacancies~\cite{smallenburg:2012,marechal:2012}, and binary mixtures~\cite{kranendonk:1991a,kranendonk:1991b}.

Simulation techniques for locating first-order transitions and critical points (CP) are fairly well established~\cite{berg:1992,panagiotopoulos:2000,binder:1981,newman:1999},
but there exists a key problem with simulating cluster crystals. Once $N_{\mathrm{c}}$ lattice sites are initialized for a $N$-particle system, the average lattice site occupancy $n_{\mathrm{c}} \equiv  N/N_{\mathrm{c}}$ cannot generally relax to its equilibrium value.  The recently developed $[N]pT$ ensemble method successfully surmounts this problem by allowing both particle number and lattice spacing to fluctuate~\cite{zhang:2012}. At temperature $T$ and pressure $p$, the equilibrium occupancy then coincides with the minimization of the constrained Gibbs free energy $G_{\mathrm{c}}(N,p,T,N_{\mathrm{c}}) \equiv \mu N +\mu_{\mathrm{c}} N_{\mathrm{c}}$, whose exact differential form~\cite{mladek:2007}
\begin{equation}
dG_{\mathrm{c}} = -SdT+Vdp+\mu dN + \mu_{\mathrm{c}}dN_{\mathrm{c}},
\end{equation}
depends not only on the entropy $S$,  the volume $V$, the chemical potential of the particles $\mu$, but also on $\mu_{\mathrm{c}}$,  the chemical potential-like quantity conjugate to $N_{\mathrm{c}}$.
If $N_{\mathrm{c}}$ is fixed, as it is in simulations of a reasonable size, the per particle free energy $g_{\mathrm{c}}(p,T,n_{\mathrm{c}})\equiv G_{\mathrm{c}}/N$ must thus be minimized to identify the equilibrium state at constant $T$ and $p$.

\begin{figure}
\includegraphics[width =\columnwidth]{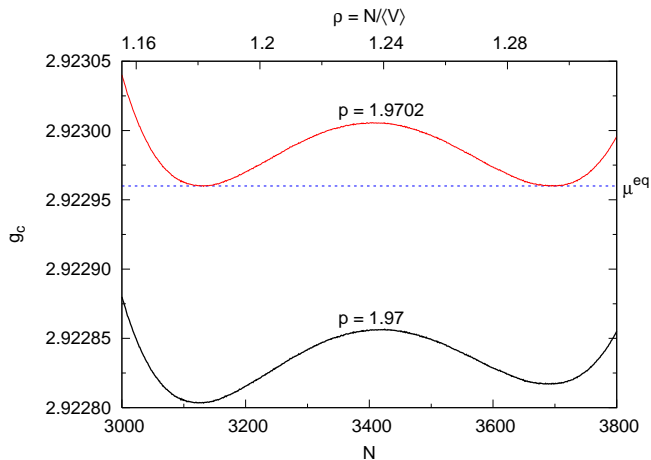} 
\caption{(Color online)  Converged $g_{\mathrm{c}}$ (black (thick) solid line) from $[N]pT$ GEM-4 simulations with $N_{\mathrm{c}}=1372$ at $T=0.045$ and $p=1.97$. After reweighting to $p=1.9702$ (red (thin) solid line), the two wells are at the same depth, which indicates that crystals at these two densities have the same chemical potential $\mu^{\mathrm{eq}}$ (dotted line). Note that the $g_{\mathrm{c}}$ has an intrinsic error of  $0.0003$, due to thermodynamic integration, which gives the pressure uncertainty of $\mathcal{O}(0.001)$ and the optimized particle number uncertainty of $\mathcal{O}(1)$.}
\label{fig:GEM4dNpT}
\end{figure}

\emph{Coexistence densities.}
Given an external field $\widetilde{G}_{\mathrm{N}}$ and a particle number window $[N_{\min},N_{\max}]$, the $[N]pT$ ensemble  samples $N$ with probability \begin{equation}
\mathcal{P}(N)\sim e^{\beta \widetilde{G}_{\mathrm{N}}} e^{-\beta G_{\mathrm{c}}},
\label{eq:PN}
\end{equation}
and iteratively determines $g_{\mathrm{c}}$ within that window~\cite{zhang:2012}. Convergence of the algorithm, which is identified  by $\mathcal{P}(N)$ being flat, provides $\widetilde{G}_{\mathrm{N}}$ differing from $G_{\mathrm{c}}$ by a constant.
This constant can then be determined by a single thermodynamic integration, neglecting the trivial thermal wavelength contribution~\cite{frenkel:2002,zhang:2012}.
Because the minimum of $g_{\mathrm{c}}$ corresponds to the equilibrium lattice occupancy, with $\mu_{\mathrm{c}}=0$, the $[N]pT$ ensemble method can also locate the coexistence
densities of an isostructural transition by adjusting the temperature and pressure so that the two minima of $g_{\mathrm{c}}$ have the same depth, i.e., the same $\mu^{\rm eq}$ (Fig.~\ref{fig:GEM4dNpT}).
If the simulated pressure differs from the coexistence pressure
at the chosen temperature,  histogram reweighting over pressure to equate the two wells is used~\cite{zhang:2012}.
By histogram reweighting over $T$ and $p$, one can also obtain coexistence densities at neighboring phase points~\cite{depablo:1999,zhang:2012}.

\begin{figure}
\begin{center}
\includegraphics[width =\columnwidth]{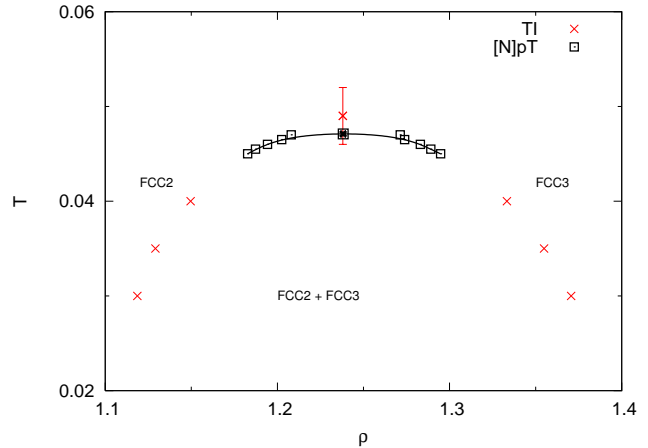}
\caption{(Color online) GEM-4 first-order FCC2-FCC3 isostructural transition. Previous thermodynamic integration (TI) results~\cite{zhang:2010b} are shown for comparison. $[N]pT$ simulations with histogram reweighting and finite-size scaling estimate $T_c(\infty) = 0.0471(2)$ and $\rho_c(\infty) = 1.2382(12)$.}
\label{fig:GEM4iso}
\end{center}
\end{figure}

We first perform $[N]pT$ simulations of  a system with $N_{\mathrm{c}}=1372$ at $T=0.045$ and $p=1.97$. The pair potential $\phi(r)$ is truncated at a cutoff distance of $1.7\sigma$, beyond which the potential energy is treated in an average way~\cite{frenkel:2002}. In addition to standard particle and logarithmic volume Monte Carlo (MC) moves~\cite{frenkel:2002}, particle insertions ($+$) and removals ($-$) are used with the acceptance ratio
\begin{equation}
\mathrm{acc}^{\pm}=
 \min\left\{1,\eta^{\pm} e^{\beta\Delta G^{\pm}-\beta\Delta E^{\pm}}\right\},
\end{equation}
where $\Delta G^{\pm} =\widetilde{ G}_{\mathrm{N\pm1}}-\widetilde{ G}_{\mathrm{N}}$, $\eta^{+}=V/(N+1)$, $\eta^{-}= N/V$, and $\Delta E^{\pm}$ is the
energy cost of inserting/removing a particle~\cite{orkoulas:2009,zhang:2012}.
The free energy from thermodynamic integration is used as initial guess for $\widetilde{G}_{\mathrm{N}}$. For a particle window of width $\Delta N = 800$, $10^7$
MC cycles are sufficient to obtain well-averaged quantities. The resulting and reweighted $g_{\mathrm{c}}$ curves are shown in Fig.~\ref{fig:GEM4dNpT}. Although an intrinsic error in $g_{\mathrm{c}}$ of the order $\mathcal{O}(1/N)$ is
introduced by the thermodynamic integration, the minimum of the $g_{\mathrm{c}}$ curve is only affected by $\delta N \sim \mathcal{O}(1)$~\cite{zhang:2012}, which also sets the error bar on the
coexistence pressure.
With this scheme, we determine a series of coexistence densities for the FCC2-FCC3 isostructural transition close to the critical point (Fig.~\ref{fig:GEM4iso}).

\emph{Critical point.}
The accurate determination of the critical point involves a  finite-size scaling approach based on the renormalization group~\cite{wilding:1995}. At the apparent critical temperature $T_c(L)$ of a finite system of linear size $L$, the distribution of the order parameter adopts a universal form~\cite{binder:1981,hilfer:1995}, once properly rescaled, that allows to extrapolate $T_c$ to the thermodynamic, i.e., infinite system size, limit. Note that under this definition, the distribution of order parameter still exhibits a double peak at the apparent critical temperature. In order to locate the GEM-4 FCC2-FCC3 isostructural critical point, we thus need to calculate the distribution $\mathcal{P}(\rho)$ of  density $\rho\equiv N/V$. We show below that one can obtain $\mathcal{P}(\rho)$  from   $G_{\mathrm{c}} = g_{\mathrm{c}} N$ by carefully considering the statistical mechanics of cluster crystals.

At fixed $N$, $p$, and $T$ ($\beta\equiv1/k_BT$, where $k_B$ is the Boltzmann constant), the constrained Gibbs free energy  $G_{\mathrm{c}}(N_{\mathrm{c}})=G_{\mathrm{c}}(N,p,T,N_{\mathrm{c}})$ as a function of $N_{\mathrm{c}}$ is effectively a Landau free energy with order parameter $N_{\mathrm{c}}$
\begin{equation}
e^{-\beta G_{\mathrm{c}}(N_{\mathrm{c}})} = \sum_{\nu} e^{-\beta G_{\nu}} \delta(N_{\mathrm{c}}^{\nu} -N_{\mathrm{c}}),
\end{equation}
where $\sum_{\nu}$ is a  sum over all microstates indexed by $\nu$ subject to the constant $NpT$ constraint. The Kronecker delta function  $\delta(x)$ is unity when $x=0$ and zero otherwise. The equilibrium Gibbs free energy $G(N,p,T)$ therefore satisfies
\begin{equation}
e^{-\beta G} = \sum_{N_{\mathrm{c}}} e^{-\beta G_{\mathrm{c}}(N_{\mathrm{c}})}\simeq e^{-\beta G_{\mathrm{c}}(N_{\mathrm{c}}^{\mathrm{eq}})},
\end{equation}
where the last approximation holds when the term with equilibrium $N_{\mathrm{c}}^{\mathrm{eq}}$ dominates the sum. Analogously, we can define Landau free energy $G_V(V)$ and $\widetilde{G}(N_{\mathrm{c}},V)$
\begin{equation}
e^{-\beta G_{V}(V)} = \sum_{\nu} e^{-\beta G_{\nu}} \delta(V_{\nu} -V),
\end{equation}
and
\begin{equation}
e^{-\beta \widetilde{G}(N_{\mathrm{c}},V)} = \sum_{\nu} e^{-\beta G_{\nu}} \delta(N_{\mathrm{c}}^{\nu} -N_{\mathrm{c}})\delta(V_{\nu} -V),
\label{eq:landauNcV}
\end{equation}
which are related by
\begin{equation}
e^{-\beta G_{V}(V)} =  \sum_{N_{\mathrm{c}}} e^{-\beta \widetilde{G}(N_{\mathrm{c}},V)}  \simeq  e^{-\beta \widetilde{G}(N_{\mathrm{c}}^V,V)}.
\end{equation}
Here again, for a given volume $V$, there exists a specific $N_{\mathrm{c}}^V$, such that $e^{-\beta \widetilde{G}(N_{\mathrm{c}}^V,V)}$ dominates the sum over
$N_{\mathrm{c}}$. For a $N$-particle system, density fluctuations result from changes in both $V$ and $N_{\mathrm{c}}$. Yet because a single specific
$N_{\mathrm{c}}^V$ dominates each $V$, we can consider that the density fluctuations are essentially along  the $N_{\mathrm{c}} =N_{\mathrm{c}}^V$ contour
of the two-dimensional $\widetilde{G}(N_{\mathrm{c}},V)$ surface. In other words, the density fluctuation is governed by
the Landau free energy $G_{V}(V)\simeq\widetilde{G}(N_{\mathrm{c}}^V,V)$ for fixed $NpT$. 
\begin{figure}
\includegraphics[width =\columnwidth]{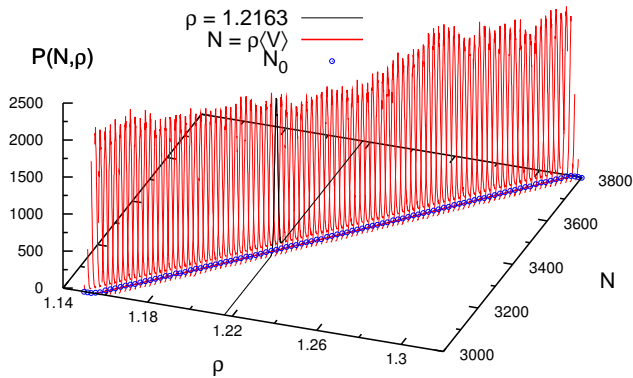}
\caption{(Color online) The sampling distribution $\mathcal{P}(N,\rho)$ at $T = 0.045$ and $p=1.97$ peaks at $N_0$ for each density. On the $\rho$-$N$ plane, the most sampled $N_0$ (blue circles) is approximately equal to $N=\rho \langle V\rangle$ (red (thick) solid line).}
\label{fig:GEM4PNrho}
\end{figure}

In  $[N]pT$ ensemble simulations, $N_{\mathrm{c}}$ is fixed while $N$ fluctuates. The density distribution is then governed by a Landau free energy $\overline{G}(N_0(\rho),\rho)$ similarly defined as in Eq.~\ref{eq:landauNcV},
where $N_0$ is the most probable particle number for a given density $\rho$. If we consider the $[N]pT$ ensemble simulation to be a series
of $NpT$ simulations, then for each $N\in [N_{\mathrm{min}}, N_{\mathrm{max}}]$ there exists an average volume $\langle V \rangle$. We can
show that if densities are binned under the definition $\rho = N/\langle V \rangle$ for each $N$, then $N\approx N_0$ for that density,
with a relative error of the order $\mathcal{O}(1/N)$. In the $[N]pT$ ensemble, the conditional probability of observing a particle number $N$ given the density $\rho$, $\mathcal{P}(N|\rho) \sim \mathcal{P}(N,\rho) \sim N \mathcal{P}(N,V)$, where the joint distribution
$\mathcal{P}(N,V)\sim e^{\beta G_{\mathrm{c}}(N,p,T,N_{\mathrm{c}})} e^{-\beta p V} e^{-\beta F_{\mathrm{c}}(N,V,T,N_{\mathrm{c}})}$~\cite{zhang:2012}.
We can thus formally write
\begin{equation}
\mathcal{P}(N,V)\sim \frac{e^{-\beta (p V+  F_{\mathrm{c}}(N,V,T,N_{\mathrm{c}}))}}{e^{-\beta G_{\mathrm{c}}(N,p,T,N_{\mathrm{c}})}} = \frac{e^{-\beta \widetilde{G}(N_{\mathrm{c}},V)}}{e^{-\beta \widetilde{G}(N_{\mathrm{c}},\langle V\rangle)}}.
\end{equation}
Given  $\rho_1$, there exists  $N_1$ such that $\rho_1=N_1/\langle V_1\rangle$. The relative probability of observing another $N_2=\rho_2 \langle V_2\rangle$ at $\rho_1$ is then
\begin{equation}
\frac{\mathcal{P}(N_2,\rho_1)}{\mathcal{P}(N_1,\rho_1)}= \frac{N_2}{N_1}\frac{\mathcal{P}(N_2,N_2/\rho_1)}{\mathcal{P}(N_1,\langle V_1\rangle)} \simeq
\frac{e^{-\beta \widetilde{G}(N_{\mathrm{c}},N_2/\rho_1)}}{e^{-\beta \widetilde{G}(N_{\mathrm{c}},\langle V_2\rangle)}}\ll 1
\end{equation}
if the linear increase in probability due to the factor $N_2/N_1$ is negligible compared to the exponential suppression. This condition is here obeyed because the standard deviation of $N_2$ from $N_1$ is $\mathcal{O}(1)$ (Fig.~\ref{fig:GEM4PNrho}). The most probable particle number $N_0$ for a given density $\rho=N/V$ is thus approximately the particle number under the definition $N = \rho \langle V \rangle$, within a relative error of the order $\mathcal{O}(1/N)$, as checked numerically in Fig.~\ref{fig:GEM4PNrho}. We can therefore use the Landau free energy $\overline{G}(N,\rho=N/\langle V \rangle)=G_{\mathrm{c}}(N=\rho \langle V \rangle,p,T,N_{\mathrm{c}})$ to analyze the density fluctuations.

Another challenge with  $[N]pT$ simulations is that both $N$ and $V$ are allowed to fluctuate, resulting in an exponential growth of the probability distribution $ \mathcal{P}(\rho = N/\langle V \rangle) \sim  e^{-\beta G_{\mathrm{c}}(N=\rho \langle V \rangle,p,T,N_{\mathrm{c}})}$, because of the increase in number of states with $N$. Although $g_{\mathrm{c}} = G_{\mathrm{c}} /N$ shows a double well, the shape of $G_{\mathrm{c}}$ is dominated by the linear increase with $N$, and the wells are nearly invisible on that scale. We can correct for this trivial exponential growth by normalizing the distribution with $e^{-\beta \mu^{\mathrm{eq}} N}$, where $\mu^{\mathrm{eq}}$ is the minimum of $g_{\mathrm{c}}$ and thus the coexistence chemical potential.  For a given system size $L$, the distribution of density is therefore $\mathcal{P}_L(\rho = N/\langle V \rangle) \sim e^{\beta (\mu^{\mathrm{eq}} N - G_{\mathrm{c}}) }$, where $G_{\mathrm{c}} = g_{\mathrm{c}} N$ is the converged field in the $[N]pT$ simulation. This approach is formally equivalent to running $[N]pT$ simulation with  $\widetilde{G}_{\mathrm{N}}= \mu^{\mathrm{eq}} N$ without iterative updating (see Eq.~\ref{eq:PN}). This alternative strategy  is, however, numerically inefficient if $\mu^{\mathrm{eq}}$ is not known in advance or if the free energy barrier between the two minima is high and the sampling efficiency is low. This simulation approach can thus be seen as one where the order parameter fluctuates at constant coexistence temperature, pressure and chemical potential.

The linear system size $L$ truncates the correlation length in a finite simulation box. In $d$-dimensional constant volume simulations, $L$ is unambiguously defined as $V^{1/d}$. When $V$ fluctuates, such as in constant $NpT$ simulations, the fixed extensive quantity $N$ is used as a measure of length scale $L\propto N^{1/d}$~\cite{wilding:1996}, but in  $[N]pT$ simulations, neither $N$ nor $V$ are fixed. As mentioned earlier, the $[N]pT$ simulation can be thought of as a series $NpT$ simulations at various $N$'s.
One may thus be tempted to propose  $N^{1/d}$ as the linear size for each $\rho = N/\langle V \rangle$. Yet the resulting collapse is not good, because in cluster crystals the system size does not straightforwardly scale with $N$, due to clustering at fixed $N_{\mathrm{c}}$. We instead make an ansatz that $L \propto N_{\mathrm{c}}^{1/d}$. 
\begin{figure}
\includegraphics[width =\columnwidth]{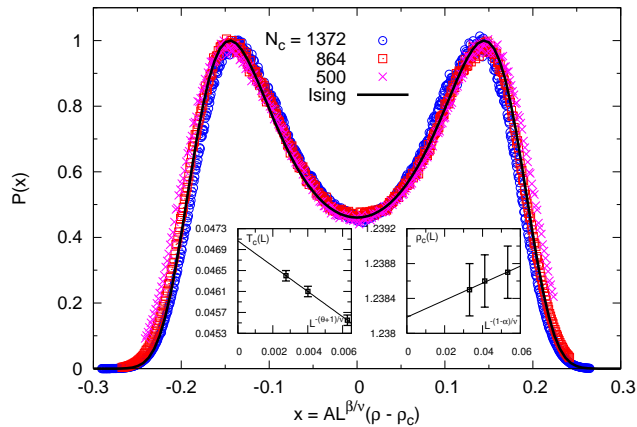}
\caption{(Color online) Finite-size scaling of the GEM-4 isostructural transition. The density distribution with appropriate critical values (Table~\ref{table:Tcrhoc})  for the system of linear size $L \propto N_{\mathrm{c}}^{1/3}$, with $N_{\mathrm{c}}=500,864, and 1372$, collapses onto the Ising universality function ($\beta/\nu = 0.518$)~\cite{tsypin:2000}. Insets: the apparent critical temperature (left) and density (right) extrapolate to the infinite system size limit (Eq.~\ref{eq:GEM4Tcscaling}).}
\label{fig:GEM4scaling}
\end{figure}

\emph{Finite-size scaling simulations.}
We perform $[N]pT$ simulations for systems with $N_{\mathrm{c}}=500,864, and 1372$ at temperatures close to their $T_c(L)$'s. For different system sizes $L\propto N_{\mathrm{c}}^{1/3}$, the distributions $\mathcal{P}_L(\rho=N/\langle V\rangle) \sim e^{\beta (\mu^{\mathrm{eq}} N - G_{\mathrm{c}})} $ at their apparent critical temperature $T_c(L)$ as a function of the scaling variable, $x=A L^{\beta/\nu}(\rho-\rho_c(L))$, collapse onto a universal function $\mathcal{P}(x)$, if the correct critical parameters $T_c(L)$, $\rho_c(L)$ and $\beta/\nu$ are chosen~\cite{wilding:1995}. We use the Ising universality exponent $\beta/\nu=0.518$ and the corresponding distribution $\mathcal{P}(x)$~\cite{tsypin:2000}. At $T_c(L)$, the ratio of the peak value of $\mathcal{P}(x)$ to its value at $x=0$ is about 2.173(4)~\cite{tsypin:2000}. With histogram reweighting, we identify the system-size dependent critical conditions (Table~\ref{table:Tcrhoc}), which result in good scaling behaviors (Fig.~\ref{fig:GEM4scaling}). The nonuniversal normalization factor $A$ is arbitrarily chosen as unity for the GEM-4 distributions, but other normalization conventions are also used in the literature~\cite{wilding:1995}. Obtaining consistent data collapse using the empirical formula in Ref.~\cite{tsypin:2000} takes $A=0.132$ for the Ising results.  The infinite-system critical temperature and density can then be extracted from the scaling relations
\begin{equation}
\begin{split}
&T_c(L)-T_c(\infty) \sim L^{-(\theta+1)/\nu}\\
&\rho_c(L)-\rho_c(\infty) \sim L^{-(1-\alpha)/\nu},
\end{split}
\label{eq:GEM4Tcscaling}
\end{equation}
where  the Ising universality class exponents are $\theta=0.54$, $\alpha =0.11$ and $\nu=0.629$~\cite{panagiotopoulos:2000} (Table~\ref{table:Tcrhoc}).
This analysis unambiguously assigns the GEM-4 isostructural transitions to the Ising universality class,
and the accuracy of the critical point location is improved by an order of magnitude (Fig.~\ref{fig:GEM4iso}).
\begin{table}
\caption{(Apparent) critical quantities for various system sizes $L\propto N_{\mathrm{c}}^{1/3}$.}
\begin{tabular}{c c c c c}
\hline
$N_{\mathrm{c}}$ & $T_c(L)$ & $\rho_c(L)$ & $p_c(L)$ & $\mu^{\mathrm{eq}}(L)$ \\
\hline
500 & 0.04555(10) & 1.2387(3) & 1.971(1) & 2.9264(3) \\
864 & 0.0461(1) & 1.2386(3) & 1.972(1) & 2.9298(3)  \\
1372 & 0.0464(1) & 1.2385(3) & 1.973(1) & 2.9316(3) \\
$\infty$ &  0.0471(2) & 1.2382(12) \\
\hline
\end{tabular}
\label{table:Tcrhoc}
\end{table}

In the liquid-vapor transition,  which lacks the particle-hole symmetry observed in the Ising model,  density $\rho$ is not an appropriate order parameter to compare with the symmetric Ising model magnetization, so  a linearly transformed density operator $\mathcal{M}=\frac{\rho - s u}{1-sr}$ with parameters $s$ and $r$, where $u$ is the energy density, is  needed ~\cite{wilding:1995}. For the GEM-4 isostructural transition, we find the symmetry of  $\rho$ to be quite good close to $T_c$.
The relatively small density gap between the FCC2 and FCC3 phases ($10$-$20\%$), compared to the orders of magnitude difference between vapor and liquid supports this observation.

In summary, we connect the constrained Gibbs free energy $G_{\mathrm{c}}(N,p,T,N_{\mathrm{c}})$ with a Landau free energy $\overline{G}(N,\rho)$ to extract  information about the density fluctuations of cluster crystals in the $[N]pT$ ensemble. At fixed $p$, $T$ and $N_{\mathrm{c}}$, the system can achieve a given density $\rho$ by various combinations of $N$ and $V$, corresponding to lattice occupancy and spacing fluctuations. Our analysis is based on the finding that a particular particle number $N_0$ dominates at a given density and $N_0$ can be approximated by $N=\rho \langle V\rangle$. If the distribution $\mathcal{P}(N,\rho)$ shown in Fig.~\ref{fig:GEM4PNrho} were to have broader peak over $N$, the lattice occupancy and thus density fluctuation would become stronger. In that case, one would need to perform random sampling in the $[N]pT$ ensemble under a two-dimensional field $\widetilde{G}(N,V)$ to identify the minimum free energy contour, but the rest of the analysis would be similar as above. It would also possible to sample the lattice occupancy fluctuation by adapting the phase switch method~\cite{bruce:1997} to the simulation of systems with fluctuating $N_{\mathrm{c}}$~\cite{wilding:2012}. This approach, however, constrains particle occupancy fluctuations -- only planes of sites can be added or removed-- which limits its accuracy, especially close to the critical point.

The method and formalism above could also be applied to the study of binary mixtures of similar components, such as hard spheres whose diameter ratio $\sigma_1/\sigma_2\lesssim 1.2$. In their crystal form, these hard sphere mixtures exhibit first-order demixing between two FCC solids above a critical pressure~\cite{kranendonk:1991a,kranendonk:1991b,hopkins:2012}. Although liquid-liquid criticality can be studied using standard MC techniques~\cite{wilding:2003}, the fixed number of lattice sites in binary solids results in a problem similar to that of cluster crystals, where both the mole fraction of each component and the lattice spacing must allowed to fluctuate. Particle insertion/removal moves can be straightforwardly replaced by particle identity changes under the Gibbs free energy per particle $g(x)=\widetilde{G}_{\mathrm{N}}/N=x\mu_1 + (1-x) \mu_2$ as a function of mole fraction $x$, with the chemical potential of the two species $\mu_1$ and $\mu_2$ mapped onto $\mu$ and $\mu_c$. Given $T$ and $p$, the coexistence is identified by the common tangent construction of the double-well $g(x)$ curve.  At the moment, the key difficulty in studying these systems is  the low acceptance of growing small particles into large ones if the simulations are not sufficiently close to the critical point, but an improved initial guess for $\widetilde{G}_{\mathrm{N}}$ could help solve this technical difficulty.

\begin{acknowledgments}
We thank  N.~Wilding for helpful discussions about the phase switch method. We acknowledge funding from National Science Foundation (NSF) Grant No. NSF DMR-1055586.
\end{acknowledgments}


\end{document}